\documentclass[12pt, dvipsnames, a4paper]{elsarticle}

\usepackage{amssymb}
\usepackage{todonotes}
\usepackage{float}
\usepackage{listings}
\usepackage{hyperref}

\lstdefinestyle{mystyle}{
    basicstyle        = \ttfamily\footnotesize,
    breakatwhitespace = false,
    breaklines        = true,
    captionpos        = b,
    keepspaces        = true,
    numbersep         = 5pt,
    showspaces        = false,
    showstringspaces  = false,
    showtabs          = false,
    keywordstyle      = \color{blue},
    stringstyle       = \color{ForestGreen},
    commentstyle      = \color{red}\ttfamily
}

\restylefloat{table}
\journal{SoftwareX}

\begin{document}

\begin{frontmatter}

\title{Omnisolver: an extensible interface to Ising spin--glass and QUBO solvers}

\author[1]{Konrad Ja\l{}owiecki}
\author[1]{{\L}ukasz Pawela}
\address[1]{Institute of Theoretical and Applied Informatics, Polish Academy of Sciences, Ba{\l}tycka 5, 44-100 Gliwice, Poland}

\begin{abstract}
We introduce a new framework for implementing Binary Quadratic Model (BQM)
solvers called Omnisolver. The framework provides an out-of-the-box dynamically
built command-line interface as well as an input/output system, thus heavily
reducing the effort required for implementing new algorithms for solving BQMs.
The proposed software should be of benefit for researchers focusing on quantum
annealers or discrete optimization algorithms as well as groups utilizing
discrete optimization as a part of their daily work. We demonstrate the ease of
use of the proposed software by presenting a step-by-step, concise
implementation of an example plugin.
\end{abstract}


\end{frontmatter}

\section*{Required Metadata}

\begin{table}[H]
\begin{tabular}{|l|p{6.5cm}|p{6.5cm}|}
\hline
\textbf{Nr.} & \textbf{Code metadata description} & \textbf{Please fill in this column} \\
\hline
C1 & Current code version & 0.0.3 \\
\hline
C2 & Permanent link to code/repository used for this code version &
\url{https://github.com/euro-hpc-pl/omnisolver} \\
\hline
C4 & Legal Code License   & Apache 2.0 \\
\hline
C5 & Code versioning system used & git \\
\hline
C6 & Software code languages, tools, and services used & Python, YAML \\
\hline
C7 & Compilation requirements, operating environments \& dependencies &
Linux, \texttt{Python >= 3.9}, \texttt{dimod}, \texttt{pluggy},  plugins may require additional
Python packages and build dependencies \\
\hline
C8 & If available Link to developer documentation/manual & \url{https://omnisolver.readthedocs.io/} \\
\hline
C9 & Support email for questions & \verb|lpawela@iitis.pl|\\
\hline
\end{tabular}
\caption{Code metadata}
\label{}
\end{table}


\section{Motivation and significance}
\label{}
The rapidly developing field of quantum information brings us ever closer to
developing practical quantum computers. Currently, we are living in an era
marked by the so-called Noisy Intermediate-Scale Quantum
(NISQ)~\cite{preskill2018quantum} devices. It comes as no surprise that these
machines have attracted attention from both the scientific and business
communities. This attention results in a myriad of proposed potential
applications for NISQ devices. Implementing these applications requires
developing appropriate software, which comes with its own set of challenges.
One of such potential obstacles, which this work aims to remedy, is the cost
of testing on actual NISQ devices.

The main way to cut quantum infrastructure access costs is to utilize frameworks
that simulate or approximate their behavior. Nonetheless, if we try this
approach, we quickly run into another problem: there are multiple frameworks
available, each of which can be based on different numerical algorithms. Here,
we are concerned with one particular NISQ architecture -- the D-Wave annealer.
The behavior of this machine can be simulated using tensor
networks~\cite{schollwock2011density,orus2014practical}, approximated using
dynamical systems approach~\cite{tiunov2019annealing}, neural
networks~\cite{wu2019solving} and, for sufficiently small problem sizes, through
a brute-force approach~\cite{jalowiecki2021brute}.

The downside of this multitude of options is that the implementations of each of
these approaches have incompatible input/output APIs. In this work, we introduce
the Omnisolver package which provides an extensible, unified API for
Ising spin-glass solvers and the actual D-Wave annealer.

The Omnisolver's core motivation lies in addressing the fragmentation and
compatibility issues that can often stymie the development of quantum software.
By providing a unified API for Ising spin-glass solvers and the D-Wave annealer,
our solution presents a uniform interface to the users, allowing the development
of algorithms to be performed in a more simplified, streamlined manner.

One of the significant advantages of this approach is that it eliminates the
need for developers to learn and adapt to the varied input/output APIs of
different solvers or to become entwined in the technical complexities of
switching between them. Instead, they can devote their full attention and
resources to designing, implementing, and fine-tuning their algorithms, hence
accelerating the overall development process. The ease of integration further
encourages experimentation with different approaches, fostering innovation
within the field.

Omnisolver is builds upon the D-Wave's dimod framework. Our software extends
dimod's capabilities in two ways. Firstly, Omnisolver's plugin library
provides additional, high-performance algorithms readily usable from Python
scripts or command line. Secondly, and more importantly, the Omnisolver's core
library provides an automatic construction of a command line interface and
a unified input/output system automatically wired to the implemented solvers.
Having a framework that not only consolidates existing methods but also
allows for the easy addition and integration of new ones is in our opinion
a valuable asset.

In this context, the claim about a reduction in the effort required to write new
algorithms is not meant to indicate that Omnisolver directly simplifies the
algorithm-writing process, but rather that it facilitates and expedites the
process by providing a more consistent, comprehensive, and accommodating
environment for developers. By simplifying the integration with various solvers
and enabling easy swapping of them for testing and comparison, Omnisolver
significantly reduces the peripheral tasks associated with quantum software
development, leaving developers more time and energy to focus on their core
task: creating and optimizing algorithms.

Compared to other similarly looking libraries like pyQUBO or QUBOTools.jl our
package provides a significant new feature.
pyQUBO~\cite{zaman2021pyqubo,tanahashi2019application} provides a platform to
formulate and solve QUBO and Ising problems. It does this by converting
combinatorial optimization problems to QUBO or Ising problems, which can then be
solved by annealing machines. While pyQUBO is a valuable tool, its scope is
limited to problem formulation and conversion; it does not provide an integrated
approach to using various solving methods.

In contrast, Omnisolver's unified API allows it to handle a broader set of
tasks. By providing a consistent interface to multiple solvers, including both
Ising spin-glass solvers and D-Wave annealer, it significantly simplifies the
process of experimenting with different approaches and offers greater
flexibility to the user.

QUBO.jl~\cite{xavier2023qubo}, a Julia library, shares a similar purpose with
pyQUBO. It assists in the formulation of QUBO problems from a combinatorial
optimization standpoint, offering various utility functions for working with
QUBO matrices. The use of Julia allows for high-performance computations,
particularly beneficial when dealing with large-scale problems. Compared to
Omnisolver, QUBO.jl provides similar problem formulation capabilities and offers
a similar, extensible platform for integrating and testing various QUBO and
Ising solvers. The main difference between QUBO.jl and Omnisolver is the
underlying technology. Omnisolver, beeing written in pure Python, allows for
seamless integration with external solvers accessed via public APIs, like for
example D-Wave's quantum annealer. Additionally, we provide a specialized plugin
for brute-force solving of QUBO problems, which guarantees finding the optimal
solution and has a reasonable execution time for instances of up to 50 variables
which allows for certification of results obtained from other methods. These
features make Omnisolver a more comprehensive tool, catering to a broader
spectrum of the QUBO and Ising problem-solving workflows.

Furthermore, it is important to acknowledge the vast landscape of Ising
machines, as reviewed in~\cite{mohseni2022ising}. These machines, both software
and hardware-based, have made significant strides in solving complex
combinatorial problems. However, many of these implementations operate in
isolation and come with their own specific input/output requirements, adding to
the complexity of algorithm development and testing.

\section{Software description}
\label{}

\subsection{Software Architecture}
\label{}

\begin{figure}
    \centering
    \includegraphics[width=\textwidth]{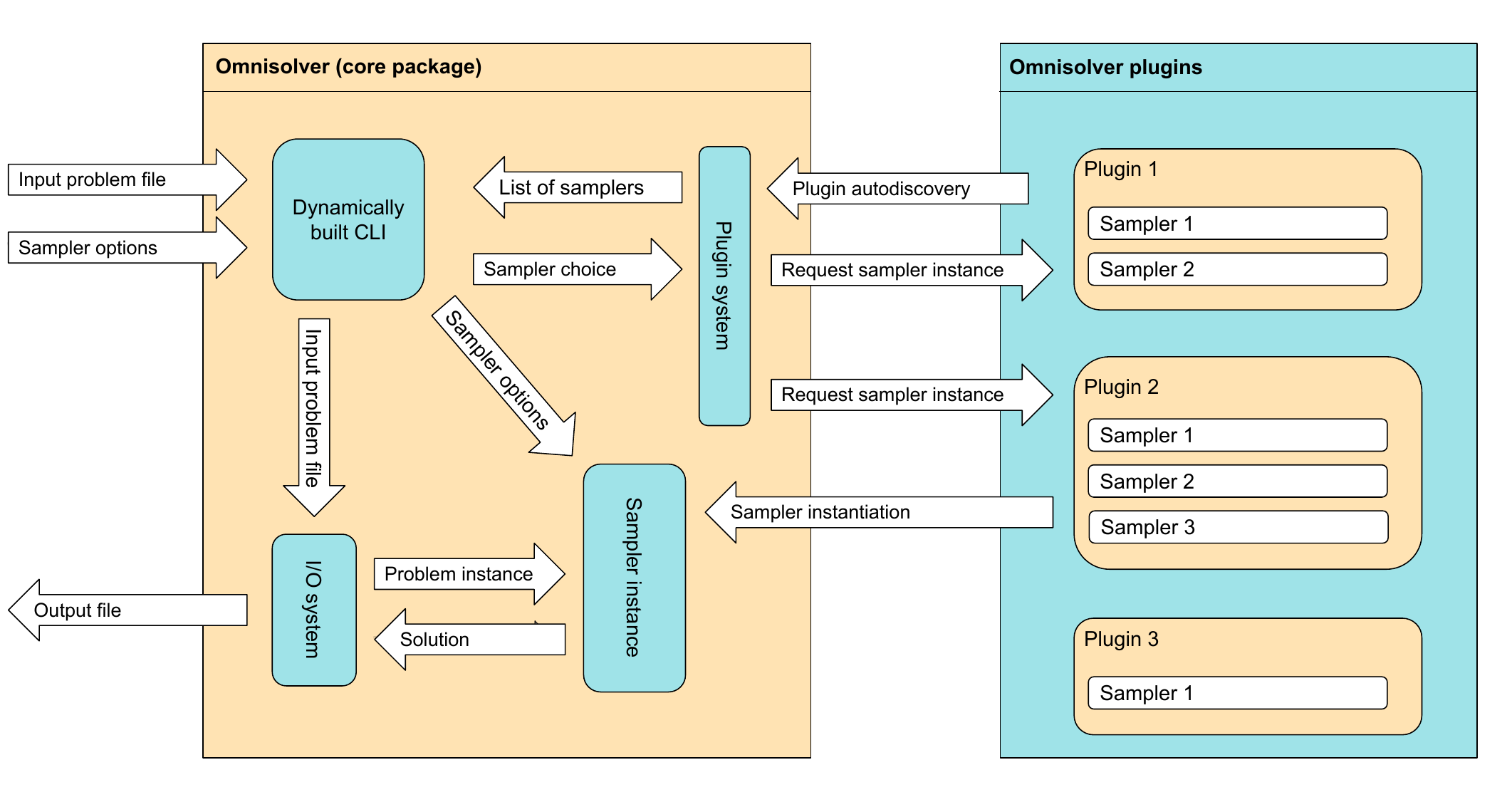}
    \caption{Omnisolver's architecture diagram.}
    \label{fig:architecture}
\end{figure}

Omnisolver is built in a modular fashion and comprises the following elements:
\begin{itemize}
    \item core Omnisolver package,
    \item plugins providing samplers built on top of \texttt{dimod} library \cite{dimod}.
\end{itemize}
The architecture of the Omnisolver, as well as the typical execution flow, is
presented in Fig.~\ref{fig:architecture}. The core package handles input/output
operations, including parsing the command line arguments, reading an input
problem file, and outputting solutions computed by samplers. It can be extended
via a plugin system built using the \texttt{pluggy} \cite{pluggy} library. The
plugins are responsible for implementing algorithms solving instances of Ising spin--glass or QUBO models (collectively known as Binary Quadratic
Models) and providing descriptions of available parameters. At the start of the
program the core package auto--discovers the plugins, and collects the
description of the samplers they implement. Based on those descriptions, a
Command Line Interface (CLI) is constructed and exposed to the user. Using the
CLI, the user selects which sampler, and with what parameters, should be used.
The core package then asks the plugin to instantiate the sampler, uses it to
solve the provided problem instance, and then serializes the output.

The plugin definition comprises:
\begin{itemize}
    \item \texttt{name} of the plugin,
    \item the \texttt{create\_sampler} function used to construct new sampler
    instances,
    \item the \texttt{populate\_parser} callback used for defining command line
    arguments used for creating sampler and running its \texttt{sample} method.
    \item \texttt{init\_args} and \texttt{sample\_args} determining which
    command line arguments should be passed respectively to sampler's
    initializer and \texttt{sample} method.
\end{itemize}
Omnisolver extensions should provide one or more callables decorated with
\texttt{omnisolver.plugin.plugin\_impl} returning an instance of \texttt{Plugin}
class.

\subsection{Software Functionalities}
\label{}

The major functionality of the Omnisolver project is providing a framework for
implementing arbitrary solvers of classical Ising spin--glasses. To this end,
the core \texttt{omnisolver} package provides:

\begin{itemize}
    \item plugin system for registering solvers inheriting from
      \texttt{dimod.Sampler},
    \item helper functions for implementing most typical plugins which remove
      the burden of writing plugin boilerplate from the end--user,
    \item input/output system automatically wired to the plugins. Currently,
      this allows reading spin--glass instances from coordinate format and
      writing solutions as a comma--separated value (CSV) file,
    \item dynamically generated command line interface (CLI). The CLI
      takes into account information provided by the registered plugins, and
      hence is able to correctly recognize parameters that need to be passed to
      each solver.
\end{itemize}

Apart from the core \texttt{omnisolver} package, we also implemented several
plugins, including:

\begin{itemize}
    \item \texttt{omnisolver-pt} implementing the parallel-tempering
      algorithm, available at \texttt{https://github.com/euro-hpc-pl/omnisolver-pt}
    \item \texttt{omnisolver-bruteforce} implementing the exhaustive
      search which can optionally be accelerated using CUDA--enabled GPU, available
      at \texttt{https://github.com/euro-hpc-pl/omnisolver-bruteforce}.
\end{itemize}

\subsection{Sample code snippets analysis}
\label{subsec:code-snippets}

\begin{figure}[!h]
  \centering
  \includegraphics[width=\textwidth]{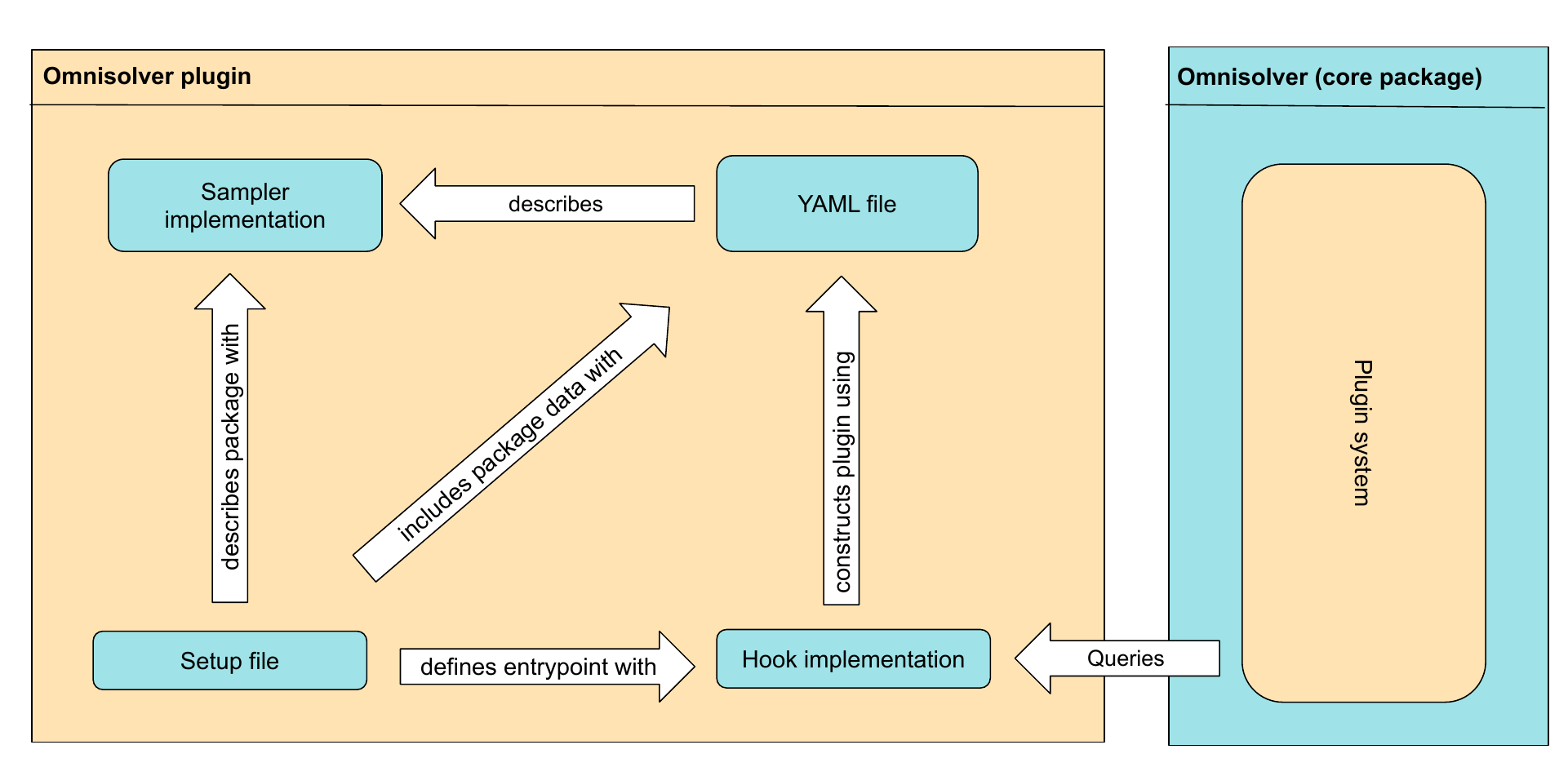}
  \caption{Omnisolver's plugin architecture diagram}
  \label{fig:plugin}
\end{figure}

Code snippets in this section demonstrate an example implementation of an
Omnisolver plugin, providing a dummy solver returning random solutions. While
this example is clearly artificial, it should be easy for the reader to extend
it to an implementation of an arbitrary nontrivial solver. The plugin is
implemented as a Python package called \texttt{dummysolver}. The layout of the
package is shown in the Listing \ref{lst:package-layout}, and the Fig.
\ref{fig:plugin} summarizes the relationship between the package's essential
components.

\begin{lstlisting}[
    caption=Directory structure of dummy omnisolver plugin,
    label={lst:package-layout}
]
+-- dummysolver
|   |-- dummy.yml
|   |-- __init__.py
|   +-- solver.py
|-- MANIFEST.in
|-- setup.cfg
+-- setup.py
\end{lstlisting}

The \texttt{dummysolver/solver.py} file, presented in Listing \ref{lst:solver},
contains the implementation of the solver. The \texttt{DummySolver} class
inherits from \texttt{dimod.Sampler} and implements its abstract \texttt{sample}
method. This method accepts an instance of \texttt{dimod.BinaryQuadraticModel}
(required by the base class' \texttt{sample} method) and an additional keyword parameter \texttt{num\_solutions} indicating how many solutions should be returned.

\begin{lstlisting}[
    language=Python,
    caption=\texttt{dummysolver/solver.py},
    label={lst:solver}
]
import dimod
import random


class DummySolver(dimod.Sampler):
    """A dummy sampler returning random solutions for given BinaryQuadraticModel."""

    @property
    def parameters(self):
        """Parameters accepted by this sampler."""
        return {"num_solutions": []}

    @property
    def properties(self):
        """Properties of this sampler."""
        return {}

    def sample(self, bqm, **parameters):
        """Uniformly sample given BQM."""
        num_samples = parameters.pop("num_solutions", 1)
        assert (
            not parameters
        ), f"Unrecognized parameters: {parameters}"

        possible_values = tuple(bqm.vartype.value)
        solutions = [
            [
                random.choice(possible_values)
                for _ in range(bqm.num_variables)
            ]
            for _ in range(num_samples)
        ]

        return dimod.SampleSet.from_samples_bqm(
            solutions, bqm
        ).aggregate()
\end{lstlisting}

To inform the Omnisolver's plugin system about \texttt{DummySolver}, one needs
to provide information about the solver's class, displayed name, and parameters
accepted by the solver and their command line counterparts. The easiest way to
achieve this is by providing the information in the YAML file, as exemplified by
Listing \ref{lst:yaml}.

\begin{lstlisting}[
    caption=\texttt{dummysolver/dummy.yml},
    label={lst:yaml}
]
schema_version: 1
name: "dummy"
sampler_class: "dummysolver.solver.DummySolver"
description: "Uniformly random dummy sampler"


init_args: []
sample_args:
  - name: "num_solutions"
    help: "number of sampled solutions"
    type: int
    default: 1
\end{lstlisting}
The \texttt{init\_args} field, specified as an empty list in the example,
defines parameters for the solver's \texttt{\_\_init\_\_} method. The
\texttt{sample\_args} list describes the parameters of the solver's
\texttt{sample} method. In principle, one could always design the plugin in such
a way that all parameters are passed to the \texttt{sample} method of the
sampler, and thus it might seem redundant to include \texttt{init\_args} as
well. However, to support samplers with a parametrized initializer that already
exist on the market, we decided to include the initializer's args as well. An
example of such a sampler is the D-Wave's \texttt{DWaveSampler}, which can
accept several parameters during its initialization \cite{ocean}.

One also needs to define the plugin's entrypoint, which in this case is located
in the package's initialization file (Listing \ref{lst:init}). Here, we used
Omnisolver's convenience function \texttt{plugin\_from\_specification} for
building the plugin from the definition read from the YAML file. Finally, the
entrypoint has to be defined in the package's setup file (Listing \ref{lst:setup})
to be picked up by the plugin system.

\begin{lstlisting}[
    language=Python,
    caption=\texttt{dummysolver/\_\_init\_\_.py},
    label={lst:init}
]
from omnisolver.plugin import (
    plugin_from_specification,
    plugin_impl,
)
from pkg_resources import resource_stream
from yaml import safe_load


@plugin_impl
def get_plugin():
    """Construct plugin from given yaml file."""
    specification = safe_load(
        resource_stream("dummysolver", "dummy.yml")
    )
    return plugin_from_specification(specification)
\end{lstlisting}

\begin{lstlisting}[
    caption=\texttt{setup.cfg},
    label={lst:setup}
]
[metadata]
name = dummysolver
description = Dummy solver plugin for omnisolver

[options]
packages = dummysolver
python_requires = >= 3.7
install_requires =
    omnisolver >= 0.0.1
include_package_data = True

[options.package_data]
dummysolver = "dummysolver.yml"

[options.entry_points]
omnisolver =
    dummy = dummysolver
\end{lstlisting}
\section{Illustrative Examples}

Examples in this section assume that \texttt{omnisolver}, \texttt{omnisolver-pt}
and \texttt{dummysolver} (presented in Subsection \ref{subsec:code-snippets})
are installed in the current Python environment.

The available solvers can be displayed by running \texttt{omnisolver -h}, as
exemplified in Listing \ref{lst:omnisolver-h}

\begin{lstlisting}[
    caption=Example of running \texttt{omnisolver -h},
    label={lst:omnisolver-h}
]
  usage: omnisolver [-h] {pt,dummy,random} ...

  optional arguments:
    -h, --help         show this help message and exit

  Solvers:
    {pt,dummy,random}
\end{lstlisting}

Help for the specific solver, automatically constructed using Python's standard
library \texttt{argparse} module, can be obtained by running \texttt{omnisolver
<solver-name> -h}. Listing \ref{lst:solver-help} shows an example output of
running command \texttt{omnisolver pt -h}.

\begin{lstlisting}[
    caption=Specific help message for the \texttt{omnisolver-pt} plugin,
    label={lst:solver-help}
]
  usage: omnisolver pt [-h] [--output OUTPUT] [--vartype {SPIN,BINARY}] [--num_replicas NUM_REPLICAS] [--num_pt_steps NUM_PT_STEPS] [--num_sweeps NUM_SWEEPS] [--beta_min BETA_MIN] [--beta_max BETA_MAX] input

  Parallel tempering sampler

  positional arguments:
    input                 Path of the input BQM file in COO format. If not specified, stdin is used.

  optional arguments:
    -h, --help            show this help message and exit
    --output OUTPUT       Path of the output file. If not specified, stdout is used.
    --vartype {SPIN,BINARY}
                          Variable type
    --num_replicas NUM_REPLICAS
                          number of replicas to simulate (default 10)
    --num_pt_steps NUM_PT_STEPS
                          number of parallel tempering steps
    --num_sweeps NUM_SWEEPS
                          number of Monte Carlo sweeps per parallel tempering step
    --beta_min BETA_MIN   inverse temperature of the hottest replica
    --beta_max BETA_MAX   inverse temperature of the coldest replica
\end{lstlisting}

Finally, let us demonstrate solving an instance of the Ising spin-glass using a
chosen solver. To this end, let us assume that the instance file
\texttt{instance.txt} as presented in Listing \ref{lst:instance} is present in
the current directory. The input file comprises rows of the form ``\texttt{i j J\_ij}''. Here, \texttt{i} and \texttt{j} are indices of variables and \texttt{J\_ij} is the
coupling coefficient between them. In the case of \texttt{i} = \texttt{j}, the
\texttt{J\_ii} coefficient is treated as linear bias (magnetic field)
acting on the \texttt{i}-th spin. For the QUBO formulation, the last
element in the row is naturally interpreted as either the quadratic or linear
term.

Running \texttt{omnisolver pt instance.txt
--vartype=SPIN} yields result similar to the one presented in Listing
\ref{lst:result}.

\begin{lstlisting}[
    caption={{
      Example file with an instance of Ising spin-glass. The Hamiltonian of the presented problem reads $H(s_{0}, s_{1}, s_{2}) = -1s_{0}s_{1} -3s_{0}s_{2}+1.5s_{1}s_{2}$.
    }},
    label={lst:instance}
]
1 2 1.5
0 1 -1
0 2 -3
\end{lstlisting}

\begin{lstlisting}[
    caption=Example output of running omnisolver on specific Ising spin-glass,
    label={lst:result}
]
0,1,2,energy,num_occurrences
1,-1,1,-3.5,1
\end{lstlisting}

\section{Impact}
\label{}

This software has two main contributions. Firstly, it provides a unified
framework for implementing Ising or QUBO solvers. As such, it should accelerate
the process of implementing new algorithms, saving the developers the time needed
for writing code handling CLI and input/output system. For the researchers, common
CLI and input/output interface should make it easier to interchangeably use a
variety of solvers, thus decreasing time needed for preparing and running
experiments. In particular, this will mainly benefit entities that face
combinatorial optimization problems in their daily routine, such as railway
companies. As an example of such an application, we refer an interested reader to
the works~\cite{domino2020quantum,domino2021quantum}.

\begin{figure}
  \includegraphics[width=\textwidth]{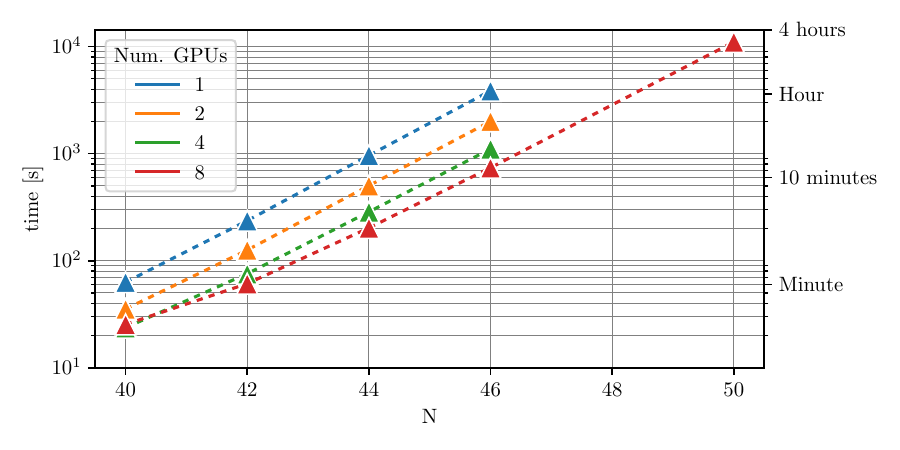}
  \caption{
    Performance of the GPU-accelerated Omnisolver-bruteforce sampler
    using a variable number of GPUs. The x-axis shows the problem size,
    and the y-axis shows the time needed to solve it using 1,2,4 and 8
    NVIDIA A100 GPUs.
  }
  \label{fig:omnisolver-bf}
\end{figure}

Secondly, Omnisolver provides several readily available samplers implemented
as its plugins, including the parallel-tempering sampler and the brute--force
sampler capable of finding low--energy spectra of instances with sizes
similar to the largest cliques embeddable on present-day D-Wave hardware.
To emphasize the usefulness of the currently implemented solver, in Fig.
\ref{fig:omnisolver-bf} we present initial performance benchmarks for Omnisolver's
bruteforce sampler.

\section{Conclusions}
\label{}

We provide the community interested in quantum computing with an
extensible framework allowing for communication with various
computational backend for solving Ising spin--glasses or QUBOs.
The plugins for the framework may implement numerical algorithms
as well as serve as interfaces to actual computing devices.

We plan further development of plugins for Omnisolver, which already
contains implementations of such algorithms as exhaustive search or
parallel tempering. The community is also encouraged
to provide their own implementations when coming up with interesting
approaches to solving Ising instances.

\section{Conflict of Interest}
No conflict of interest exists: We wish to confirm that there are no known
conflicts of interest associated with this publication and there has been no
significant financial support for this work that could have influenced its
outcome.

\section*{Acknowledgements}
\label{}

This work is supported by the project “Near-term quantum computers Challenges, optimal implementations and applications” under Grant Number POIR.04.04.00–00–17C1/18–00, which is carried out within the Team-Net programme of the Foundation for Polish Science co-financed by the European Union under the European Regional Development Fund. We gratefully acknowledge Poland’s high-performance computing infrastructure PLGrid (HPC Centers: Cyfronet Athena) for providing computer facilities and support within computational grant no. PLG/2022/015734.

\bibliographystyle{elsarticle-num}
\bibliography{omnisolver.bib}

\end{document}